\documentclass[sigconf]{acmart}

\usepackage{booktabs}       
\usepackage{amsfonts}       
\usepackage{nicefrac}       
\usepackage{microtype}      

\usepackage{balance}

\def\BibTeX{{\rm B\kern-.05em{\sc i\kern-.025em b}\kern-.08emT\kern-.1667em\lower.7ex\hbox{E}\kern-.125emX}}
    
\copyrightyear{2020}
\acmYear{2020}
\setcopyright{acmlicensed}
\acmConference[AIES '20]{Proceedings of the 2020 AAAI/ACM Conference on AI, Ethics, and Society}{February 7--8, 2020}{New York, NY, USA}
\acmBooktitle{Proceedings of the 2020 AAAI/ACM Conference on AI, Ethics, and Society (AIES '20), February 7--8, 2020, New York, NY, USA}
\acmPrice{15.00}
\acmDOI{10.1145/3375627.3375835}
\acmISBN{978-1-4503-7110-0/20/02}

\begin{document}

\fancyhead{}

\title{Defining AI in Policy versus Practice} 


\author{P. M. Krafft}
\authornote{Work conducted at the University of Washington.}
\affiliation{%
  \institution{Oxford Internet Institute\\University of Oxford}
}
\orcid{0000-0001-8570-2180}
\email{p.krafft@oii.ox.ac.uk}

\author{Meg Young}
\affiliation{%
  \institution{Information School\\University of Washington}
}
\orcid{0000-0002-9300-8575}
\email{megyoung@uw.edu}

\author{Michael Katell}
\affiliation{%
  \institution{Information School\\University of Washington}
}
\orcid{0000-0003-2200-6246}
\email{mkatell@uw.edu}

\author{Karen Huang}
\affiliation{%
  \institution{Harvard University}
}
\orcid{0000-0001-6636-6273}
\email{karenhuang@g.harvard.edu}

\author{Ghislain Bugingo}
\affiliation{%
  \institution{Information School\\University of Washington}
}

%
\renewcommand{\shortauthors}{Krafft et al.}

%
\begin{abstract}
Recent concern about harms of information technologies motivate consideration of regulatory action to forestall or constrain certain developments in the field of artificial intelligence (AI). However, definitional ambiguity hampers the possibility of conversation about this urgent topic of public concern.  Legal and regulatory interventions require agreed-upon definitions, but consensus around a definition of AI has been elusive, especially in policy conversations. With an eye towards practical working definitions and a broader understanding of positions on these issues, we survey experts and review published policy documents to examine researcher and policy-maker conceptions of AI. We find that while AI researchers favor definitions of AI that emphasize technical functionality, policy-makers instead use definitions that compare systems to human thinking and behavior.  We point out that definitions adhering closely to the functionality of AI systems are more inclusive of technologies in use today, whereas definitions that emphasize human-like capabilities are most applicable to hypothetical future technologies.  As a result of this gap, ethical and regulatory efforts may overemphasize concern about future technologies at the expense of pressing issues with existing deployed technologies.
\end{abstract}

%
%


%

%
\maketitle

\section{Introduction}


As computational systems have come to play an increasing role in making predictions about people, the downstream social consequences of artificial intelligence (AI) have garnered growing public attention. In the short term, evidence indicates that machine learning (ML) algorithms could contribute to oppression and discrimination due to historical legacies of injustice reflected in training data \cite{angwin2016machine,buolamwini2018gender,dastin2018amazon}, directly to economic inequality through job displacement \cite{duckworth2019inferring}, or through a failure to reflectively account for who benefits and who does not from the decision to use a particular system \cite{hoffmann2019WhereFairnessFails}. In the long term, some believe AI technologies pose an existential risk to humanity by altering the scope of human agency and self-determination \cite{muller2016future}, or by the creation of autonomous weapons \cite{asaro2012banning}.
For many researchers in machine learning, the answer to these challenges has been technical, and there has been a growth of work to promote computational approaches to fairness, accountability, and transparency in machine learning (e.g., \cite{blum2018preserving,kim2018fairness}).  For others, these challenges may require regulatory intervention.
In the regulatory line, widespread public concerns regarding the social impacts of AI have led a growing number of organizations to create policy and ethical recommendations for AI governance \cite{mitt2019ssrn}.

Here, we examine the regulatory approach and the extent to which current policy efforts are in meaningful conversation with existing AI and machine learning research, beginning with the simplest matter of definitions.
Given that lawmakers are not prima facie technologists or AI researchers, it is important to understand policymakers' operational definitions of AI on the path to effective governance choices.
Policymakers' understandings of AI may differ from those of AI researchers. For example, recent findings on municipal technology policy found that government employees did not define Automated License Plate Readers (ALPR) or Booking Photo Comparison Software, which rely on optical character recognition and facial recognition, as AI or machine learning systems \cite{young2019algorithmic}. 
Where policymakers do not have a clear definition of AI for regulatory purposes \cite{howe2019a} bureaucrats do not know which systems fall under new laws in the implementation phase. Conceptual clarity on this issue would empower a new generation of algorithmic oversight regulation. 

A \textit{policy-facing definition} of AI is one that facilitates policy implementation. Definitional challenges of automation persist across domains. In the case of autonomous weapons, a major barrier to consensus in international discussions has been lack of agreement over the definition of an autonomous weapon \cite{ccw2019laws,conn2016laws}. Furthermore, policy documents often use definitions of AI that are difficult to apply from the standpoint of policy implementation. For example, the AI4People Scientific Committee defines AI as ``a resource [that] can hugely enhance human agency'' \cite{floridi2018ai4people}. As another example, the IEEE Global Initiative on Ethics of Autonomous and Intelligent Systems defines AI as ``autonomous and intelligent technical systems [that] are specifically designed to reduce the necessity for human intervention in our day-to-day lives'' \cite{ieee2017ieee}. While these definitions are conceptually illuminating in highlighting the role of humans in AI, they are (i) too ambiguous to usefully apply in regulatory approaches to governance, and (ii) tend to misapprehend AI's current capabilities. Policymakers concerned about the disparate impact of ML applications thus risk overlooking currently deployed technology in favor of next-generation or speculative systems. 

This lack of a policy-facing definitions, along with a possible disconnect between policymakers' and AI researchers' conceptions of AI, motivate the current study. How do policy documents and AI researchers conceptualize AI technologies? How may these conceptualizations differ?  While previous work has hinted at answers to these questions, no systematic evaluation has been conducted.  We employ a mixed methods social scientific approach to address these questions. Drawing upon on Russell and Norvig's  typology of AI definitions \cite{russell2016artificial},  we first use this typology to analyze results from a series of surveys of researchers in AI to understand how researchers define AI. We then conduct a document analysis of major published policy documents in order to understand policymakers' understanding of AI through a close analysis of the definitions those documents use. Our use of a single well-established typology of AI definitions allows us to systematically compare disparate survey and document data sources to achieve a meaningful understanding of how published policy documents' definitions of AI compare to AI researchers' definitions of AI.

\section{Previous Work}
\begin{table}[]
\centering
\begin{tabular}{l|p{1.23in}|p{1.23in}|}
\cline{2-3}
                            &  Behaving                                                                                   & Thinking                                                                                                                                       \\ \hline
\multicolumn{1}{|l|}{Human} &  Definitions emphasizing meeting human performance or behaving like people do in particular tasks or situations     & Definitions emphasizing processing information in ways inspired by human cognition                         \\ \hline
\multicolumn{1}{|l|}{Ideal} & Definitions emphasizing functioning optimally according to a specification of optimality, such as by accomplishing specified tasks or goals & Definitions emphasizing processing information according to idealized rules of inference   \\ \hline
\end{tabular}
\caption{Russell and Norvig \cite{russell2016artificial} divided definitions of AI according to two dimensions: human versus ideal, and thinking versus behaving. All definitions of AI are classifications of computer programs as "intelligent" or not according to some characteristics. A human thinking definition of AI classifies a machine or a computer program as AI as being able to imitate human cognition, i.e. if it is able to imitate the way people think, either in a particular task or in a range of situations. In contrast, an ideal thinking definition is one that emphasizes AI  as reasoning correctly according to mathematical rules given a set of assumptions about the world, such as deductive logic or statistical inference. A human behaving definition of AI emphasizes AI as imitating the way people behave or complete particular tasks---the distinction between human thinking versus behaving is that a human behaving AI might be able to do the same thing a person does but for different reasons, such as using hard-wired rules. Finally, an ideal behaving definition emphasizes AI as correctly accomplishing the tasks or goals it is programmed to accomplish. Ideal behaving AI might ``think'' (i.e., process information) and behave in unusual or unfamiliar ways, but ways that are effective for is specified tasks or goals. }
\label{table:norvig}
\end{table}

Researchers in AI have long recognized the lack of definitional consensus in the field. According to Agre \cite{agre1997lessons}, the lack of a definition was generative: ``Curiously, the lack of a precise, universally accepted definition of AI probably has helped the field to grow, blossom, and advance.'' Even as the field has disagreed widely in practice, in their foundational textbook, ``Artificial Intelligence: A Modern Approach,'' Russell and Norvig  \cite{russell2016artificial} examine the goals of the field by considering how AI is defined in eight other textbooks published between 1978-1993, finding that definitions can be classified into four primary approaches. Namely, AI as a field is concerned with building systems that (i) think like humans, (ii) act like humans, (iii) think rationally, and (iv) act rationally. (See Table \ref{table:norvig}.) 
 However, according to Sweeney \cite{sweeney2003s}, the lack of alignment between different conceptions of AI poses a risk to the field. In order to advance a critique that the term AI suffers from ``too many different and sometimes conflicted meanings,'' Sweeney used this typology to characterize 996 AI research works cited in Russell and Norvig's textbook based on her sense that a textbook cites works representative of or important to the field. Of these, nearly all (987)  fell into one of the latter two approaches, i.e. to create artifacts that pursue rational (a.k.a. ``ideal'') thinking and behavior. Sweeney also notes large definitional shifts in the field over time, as well as low proportions of references in common between textbooks published only a few years apart. Some definitional variation may proceed from expectations of AI that are relative to the current capabilities of computing. An early AI researcher remarked that ``AI is a collective name for problems which we do not yet know how to solve properly by computer'' \cite{michie1971formation}. Similarly, John McCarthy states, ``As soon as it works, no one calls it AI anymore'' \cite{vardi2012artificial}. 
In recent years, there has been a growing amount of systematic inquiry into non-expert descriptions and understandings of AI. Awareness of AI is growing; Cave et al. \cite{cave2019scary} find that in a nationally representative survey in the United Kingdom, 85\% of respondents had heard of AI. Among a subset of 622 people who provided a description of AI, 261 referred to computers performing tasks like ``decision making,'' ``learning,'' or ``thinking.'' Another 155 respondents described it as synonymous with robots. As the authors note, ``This conflation is understandable...[but] could be problematic. Imagining AI as embodied will lend itself to some narratives more than others [such as] worries of gun-toting killer robots rather than the real-world risk of algorithmic bias'' \cite{cave2019scary}. Confirming this finding, other recent work to understand non-experts' primary concerns about AI has found that popular imaginaries draw on imagery from media \cite{fast2017long}. A focus on humanoid AI contributes to misinformed public perception and ``overshadow[s] issues that are already creating challenges today'' \cite{cave2018portrayals}.
In response to these challenges, scholars argue for changing language used about AI to sharpen its conceptual clarity. Johnson and Verdicchio \cite{johnson2017reframing} identify two primary concerns with the public reception of expert discourse on AI. First, the concept of ``autonomy'' is misapprehended as machines possessing free will and self interest. Second, a discussion highlighting the ``system'' has obscured the degree to which applications are \textit{sociotechnical}, reflecting a range of human choices, negotiations, and impacts in their design and use. To address these concerns, the authors advocate for re-framing conversations around machine autonomy to foreground human actors and the broader sociotechnical context in which such systems are embedded, an injunction forcefully argued by other critical scholars as well \cite{elish2018situating}. Other authors, such as Joanna Bryson \cite{bryson2019past}, have offered definitions that begin to synthesize these considerations with classical definitions of AI, conceived of as idealized sensing-acting machines.

\section{Methods}
The aims of this study are twofold: To describe the way that AI researchers and policymakers define AI, and to examine possible gaps between these definitions as they relate to implications for policy-making. We employed a multi-method social scientific approach. 
Our methodological motivation was complementary in combining different types of data \cite{small2011conduct}. 

We conducted two surveys of AI and ML researchers in May and July 2019.  Our first survey includes data from 98 researchers and our second includes data from 86.  We confirmed that our sample has the experience and expertise relevant to our research questions. We also assessed sample bias by confirming that the demographics of our sample match the demographics of the field of AI.  Further, we replicated both surveys with an alternative snowball sampling recruitment methodology with 108 and 44 researchers participating. Images of all the stimuli used in these surveys and results from our snowball sample replication are presented in our supplementary materials.\footnote{Full paper with supplementary materials is available online at \url{https://papers.ssrn.com/sol3/papers.cfm?abstract_id=3431304}.} 




\begin{figure}
    \centering
    \includegraphics[width=0.32\linewidth]{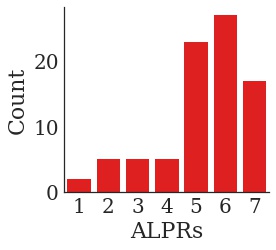}
    \includegraphics[width=0.32\linewidth]{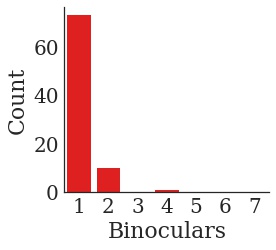}
    \includegraphics[width=0.32\linewidth]{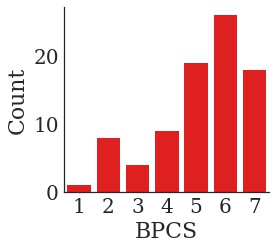}\\
    \includegraphics[width=0.32\linewidth]{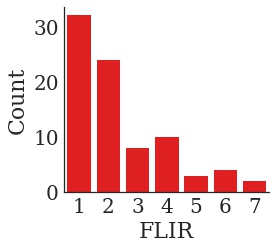}
    \includegraphics[width=0.32\linewidth]{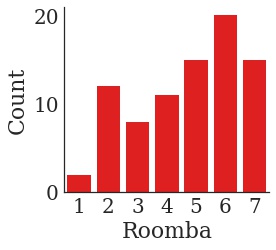}
    \includegraphics[width=0.32\linewidth]{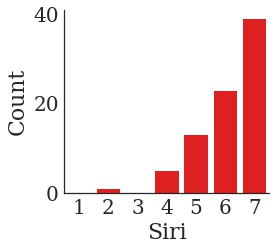}
    \caption{Histograms of AI researcher ratings of technologies as relying on AI. (1 = Strongly disagree, 7 = Strong agree)}
    \label{fig:tech-classes}
\end{figure}

\subsection{AI Researcher Survey}
We first conducted two surveys to understand AI researchers' opinions regarding definitions of AI and views on the social impacts of AI. In the first of these researcher surveys, participants began by rating the extent to which they would classify six particular technologies as AI. In choosing the six particular technologies we presented, we drew on prior work of ours that identified   different conceptions of certain surveillance technologies among city stakeholders in Seattle \cite{young2019algorithmic}.  We chose four surveillance technologies drawn from this previous study, and supplemented those four with a popular virtual assistant and a popular robot.  These six technologies were Automated License Plate Readers (ALPRs), binoculars, Booking Photo Comparison Software (BPCS), Forward-Looking Infrared camera (FLIR), Roomba, and Siri. For each technology, we presented images and descriptions, and we asked; ``Please indicate your response to the following statement: ALPRs (Automated License Plate Readers) are an artificial intelligence (A.I.)-based technology.''  We presented a 7-point Likert scale ranging from ``Strongly disagree'' to ``Strongly agree''.  We presented these technologies in randomized order. After these initial questions on classifying AI technologies, we then offered participants the option to define the term AI themselves.
Finally, participants provided their opinions regarding particular social impacts of AI. Based on our understanding of the public discourse around AI, we focused on the harms of existential risk, economic inequality, and oppression and discrimination. The format of our questions in this part of the survey was in the form ``Please indicate your response to the following statement: The potential for an existential threat to humanity is a relevant issue to consider in the research and development of A.I. technologies.''  We used the same 7-point Likert scale as in our technology classification questions.


In our second survey, rather than soliciting classifications of technologies or soliciting definitions of AI, we asked participants to directly selected the characteristics of definitions of AI they identified as most important and directly selected a favored example definition among one definition of each of the four types we used as classes of definitions. 

\subsubsection{Survey Recruitment}

We used multiple recruitment methods for our surveys. In our first survey, we successfully recruited from two major international AI mailing lists, and from AI and ML mailing lists at four U.S. universities highly ranked in AI and ML according to the U.S. News graduate ranking\footnote{https://www.usnews.com/best-graduate-schools/top-science-schools/artificial-intelligence-rankings}  and one highly ranked university in the U.K.
Survey participation was voluntary and uncompensated. 
In total, 195 people opened a survey link, and 103 participants completed the survey. We use partial data from all participants who consented.  In our analysis we subsampled our data to only include participants self-identifying as AI researchers or publishing in at least one of five AI/ML conferences we asked about (NeurIPS, ICML, ICLR, IJCAI, or AAAI). This left a final sample size of 98 AI researchers for our analyses. We collected demographic information to assess the representativeness of the data we analyzed from this survey. We also replicated our results in a snowball sample through social media and the authors' professional networks.

In our second survey we successfully recruited from two major international AI mailing lists, and one U.S. university mailing list oriented towards computer science that includes alumni and visitors. 
Survey participation was voluntary and uncompensated. 
In total, 201 people opened the survey link, and 112 participants completed the survey. We use partial data from all participants who consented.  In our analysis we again subsampled our data to only include participants self-identifying as AI researchers or publishing in at least one of five AI/ML conferences we asked about (NeurIPS, ICML, ICLR, IJCAI, or AAAI). This left a final sample size of 86 AI researchers for our analyses. We collected demographic information to assess the representativeness of the data we analyzed from this survey. We also replicated this survey in a snowball sample through social media and the authors' professional networks.

\subsubsection{Survey Population Expertise}


Of the AI researcher participants in our survey, all but one reported having at some point now or in the past considered themselves AI researchers. 68\% reported having extensive experience in AI research or development, and 15\% reported having some formal education or work experience relating to AI.  100\% reported that they ever read academic publications/scholarly papers related to AI or machine learning. 22\% reported having published at NeurIPS, 19\% at ICML,  17\% at AAAI, 13\% at IJCAI, 7\% at ICLR, and 23\% at other conferences (including UAI, CVPR, AAMAS, ICRA, IROS, and ``IEEE''). 78\% reported that they considered any of their own work or projects as ``very related'' to the development of AI, while 18\% reported their work or projects as only ``somewhat related,'' and 4\% as ``a little related.''

\subsubsection{Selection Bias in our Sample Population}
To assess the representativeness of our biased sample, we checked the distribution of self-reported demographic variables.  In our first survey, 25\% reported Female as their gender, 70\% reported Male, 1\% entered their own gender identity, and 4\% indicated they preferred not to say.  Using categories based on the U.S. Census Bureau classification system, and allowing multiple selections, our subsample of AI researchers consisted of 0\% identifying as American Indian or Alaska Native, 26\% as Asian, 1\% Black or African American, 4\% Hispanic, 0\% Native Hawaiian or Pacific Islander, and 60\% White. 4\% entered their own race or ethnicity and 5\% indicated they preferred not to say.
In our second survey, which used a different phrasing of our gender question \cite{spiel2019better}, 21\% reported Woman as their gender, 77\% Man, and 0\% Non-binary. 0\% entered their own gender identity and 2\% indicated they preferred not to say.  Using categories based on the U.S. Census Bureau classification system, and allowing multiple selections, our subsample of AI researchers consisted of 0\% identifying as American Indian or Alaska Native, 17\% as Asian, 6\% Black or African American, 5\% Hispanic, 0\% Native Hawaiian or Pacific Islander, and 61\%  White. 3\% entered their own race or ethnicity, and 8\% indicated they preferred not to say.
Authoritative statistics on the demographics of the fields of AI and machine learning were not readily available at the time of writing \cite{disc2019myers}, but existing estimates place the percentage of women publishing in AI and ML at around 15-20\%, with the percentage of women working in computing more broadly being last estimated as around 25\%~\cite{disc2019myers} .\footnote{\url{https://jfgagne.ai/talent-2019/}}\footnote{\url{https://www.wired.com/story/artificial-intelligence-researchers-gender-imbalance/}}
Estimates of racial and ethnic diversity are less reliable but place the percentage of black researchers in AI potentially as low as less than 1\% and only up to roughly 5\% working in tech companies more broadly. One estimate places the percentage of Hispanic tech workers at around the same level \cite{disc2019myers}.  We were not able to find counts of other racial or ethnic groups in the field of AI. A Pew Report shows computer workers in the U.S. as mostly White (65\%), followed by Asian (19\%), Black (7\%), and Hispanic (7\%).\footnote{\url{https://www.pewsocialtrends.org/2018/01/09/diversity-in-the-stem-workforce-varies-widely-across-jobs/}} According to these estimates, the demographics of our survey appear to be relatively representative of the community of AI researchers along these dimensions, which suggests, at least along these dimensions, the bias in our data from non-random sampling is small.

\subsection{Document Analysis of Policy Documents}
We also analyzed published policy documents in order to explore how policymakers conceptualize AI. We drew upon a comprehensive inventory of available AI ethics and policy documents published between 2017 and 2019 indexed in the prominent international non-profit research organization, Algorithm Watch's, AI Ethics Guidelines Global Inventory.\footnote{https://algorithmwatch.org/en/project/ai-ethics-guidelines-global-inventory/} The inventory contains documents authored by governments, non-governmental organizations, and firms offering guidance on AI governance. Starting with the 83 documents in the collection at the time of data collection in July 2019, we analyzed all English-language documents whose titles mentioned artificial intelligence (40 documents), removing one duplicate document from the analysis. 
Within each document, we identified definitions of AI using keyword searching, using the stemmed keywords ``defin'', ''artificial intelligence'', ''AI'', and ``A.I.''. We analyzed the definitions identified through this search using a qualitative coding procedure. Three authors independently conducted coding; one coder works in artificial intelligence, one in science and technology studies, and one in information policy and ethics.
Following Sweeney \cite{sweeney2003s}, we classify definitions being used by researchers and policy documents according to Russell and Norvig's taxonomy \cite{russell2016artificial} describing four general approaches to defining AI. 
For each of these categories, we coded the approach that was most pronounced in emphasis in the definition. 
At least two of the three coders agreed on all but two cases (95\%). 

\begin{table}[t]
\centering
\footnotesize
\begin{tabular}{|l|p{1.2in}|l|}
\hline
\textbf{Type} & \textbf{Definition}                                                                                                                                                            & \textbf{Source}        \\ \hline
Human behavior              & The science of making computers do things that require intelligence when done by humans                               & The Alan Turing Institute                     \\ \hline
Human thought               & The imitation of human intelligence processes by machines, especially computer systems.                                  & Deutsche Telekom                 \\ \hline
Ideal behavior              & Systems that display intelligent behaviour by analysing their, environment and taking actions---with some degree of autonomy---to, achieve specific goals.                                     & EU Commission                        \\ \hline
Ideal thought               & AI is computer programming that learns and adapts & Google \\ \hline
\end{tabular}
\caption{Example definitions of AI sorted by Russel \& Norvig typology.}
\label{tab:def-table}
\end{table}

\section{Results}

\subsection{AI Researcher Survey}


\begin{figure}
  \begin{center}
    \includegraphics[width=0.2\textwidth]{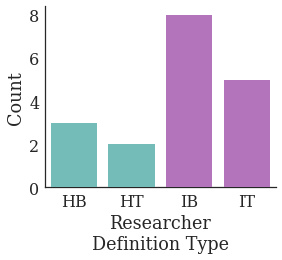}
    \includegraphics[width=0.2\textwidth]{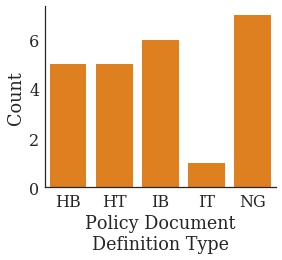}
  \end{center}
  \caption{Histogram of types of definitions AI researchers gave in our first survey, which favor ideal (rational) behavior/thinking definitions over human-imitating behavior/thinking, compared to histogram of types of definitions published AI policy document gave, which favor human-type definitions over ideal-type. Codes: I = Ideal (rational), H = Human-imitating, T = Thinking, B = Behaving (following Russell and Norvig), NG = None given.}
  \label{fig:survey-1}
\end{figure}


In our first survey, in which we asked participants to classify technologies as AI or not without providing a definition, we found a large degree of consensus in the classifications of five of the six  technologies we presented. Figure \ref{fig:tech-classes} shows that AI researchers tend to agree that ALPRs (80\% agree at least somewhat), BPCS (74\%), and Siri (93\%) are AI technologies, while binoculars (0\%) and FLIR (11\%) are not.  The one technology for which there was substantial disagreement was Roombas, with only 60\% classifying Roombas as AI. These results indicate that, at least when it comes to classifying particular technologies as AI or not, researchers often more or less agree, even in cases such as ALPRs that have been observed to be contentious for policymakers in prior research.

\begin{figure}
  \begin{center}
    \includegraphics[width=0.2\textwidth]{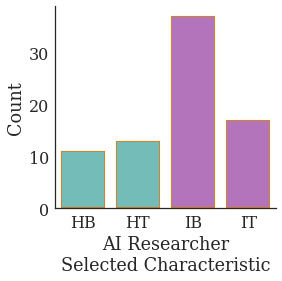}
    \includegraphics[width=0.2\textwidth]{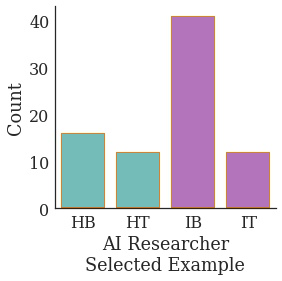}
  \end{center}
  \caption{Histogram of types of definitions AI researchers selected as favored characteristics and best example definitions in our second survey. Codes as in Figure \ref{fig:survey-1}.}
  \label{fig:survey-2}
\end{figure}

We also asked an optional question in this first survey for participants to offer their own definition of AI.  18 researchers wrote in definitions, which two authors independently coded according to Russell and Norvig's typology of human-imitating thinking (HT), human-imitating behavior (HB), ideal (rational) thinking (IT), and ideal behavior (IB).  Initial agreement across the two coders was 56\%, and the two coders came to consensus where there were disagreements.
Consistent with a 2003 analysis of types of AI in the published literature at the time \cite{sweeney2003s}, Figure \ref{fig:survey-1} shows that AI researchers tend to favor ideal thinking/behavior in their definitions of AI (72\% used ideal).  In other words,  AI researchers who provided definitions in our first survey favored definitions of AI that emphasize mathematical problem specification and system functionality over definitions that compare AI to humans.

Our second survey (results shown in Figure \ref{fig:survey-2}) was designed to provide a more direct test of this finding from the first survey.  Here we asked only for demographics and two fine-grained questions about how to define AI. We presented four example definitions that typify each definition category (HT, HB, IT, IB) and asked participants to select the best definition. We also directly asked which among the categories human-like thinking, human-like behavior, ideal/rational thinking, or ideal/rational behavior the participants judged to be most important to characterizing AI.  This simplified survey design qualitatively replicated the result of our first survey. 
We found that 65\% of AI researchers in this survey preferred ideal-type example definitions and 66\% of AI researchers selected the ideal-type categories of definitions as most important for characterizing AI.

\subsection{Policy Document Analysis}
Since the population of AI policymakers is less readily accessible than the population of AI researchers, we rely on content analysis of published policy documents rather than an interview to characterize the types of definitions of AI that tend to be used in policy-making. The documents we analyzed are published by governments, non-governmental organizations, and companies 
actively engaged in AI ethics and regulation (e.g. AI Now, EU Commission, Microsoft, Open AI, etc.). AI policy and ethics documents are the documents that policymakers consult and produce while crafting regulation and are more readily accessible than the broadly distributed set of people drafting them.  
This comparison allows us to analyze what published policy documents say about AI to what AI researchers think about AI.

We find that, in contrast to AI researchers, published policy documents tend to use definitions of AI that emphasize comparison to human thought or behavior (57\% of definitions place this emphasis). For example, the Science and Technology Committee in the UK Parliament House of Commons defines AI as ``a set of statistical tools and algorithms that combine to form, in part, intelligent software... to simulate elements of human behaviour such as learning, reasoning and classification'' \cite{housec2018ai}. Other definitions mention that AI is ``inspired by the human mind'' \cite{itic2017a}, ``capable of learning from experience, much as people do'' \cite{smith2018future}, or ``seek to simulate human traits'' \cite{singapore2019a}. Several definitions emphasize the autonomous capabilities of these systems. 

\begin{figure}
    \centering
    \includegraphics[width=0.32\linewidth]{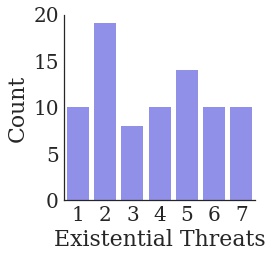}
    \includegraphics[width=0.32\linewidth]{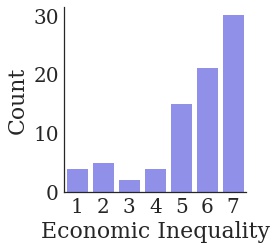}
    \includegraphics[width=0.32\linewidth]{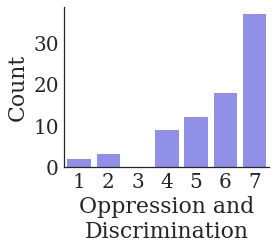}
      \caption{Histograms of AI researcher ratings of issues as relevant to AI. (1 = Strongly disagree, 7 = Strong agree)}
    \label{fig:issues}
\end{figure}

Of the 40 documents we analyzed, 15 documents (38\%) did not provide any definition of AI at all, yet issued policy recommendations about AI. This finding affirms the need for a policy-facing definition of AI that would help adhere the growing number of guidance documents to technologies relevant to their implementation. In another case, policy guidelines from the AI Now Institute at New York University relied on Russell and Norvig's \cite{russell2016artificial} typology of definitional approaches by way of a definition, defining AI as ``systems that think like humans, systems that act like humans, systems that think rationally, systems that act rationally'' \cite{whittaker2018AINow2018}.  Frequency of types encountered in our policy review appear in Figure \ref{fig:survey-1}. Example definitions sorted into the typology are shown in Table \ref{tab:def-table}.



\subsection{Quantitative Analysis}
To avoid concerns about our qualitative coding procedure, we also replicated our analysis with a quantitative method using simple natural language processing. For this quantitative analysis, we use a keyword-based classifier to judge whether definitions are using humans as a comparison point for AI simply by searching for whether either definition provided contains or does not contain the word ``human". We find that 28\% of the definitions given to us by AI researchers use the word ``human", while 62\% of definitions from published policy documents include it.

\subsection{Issue Analysis}
Given that policy documents use more human-like definitions of AI, while AI researchers favor definitions that emphasize technical problem specification and functionality, which of these classes of definitions is more relevant to concerns people express about AI?  Our first researcher survey included questions addressing this question. We analyzed what social issues AI researchers view to be relevant to AI; existential threats to humanity, economic inequality and the future of work, and oppression and discrimination (the first being commonly associated with human-like AI and the latter two with more mundane AI).  

\begin{figure}
    \centering
    \includegraphics[width=\linewidth]{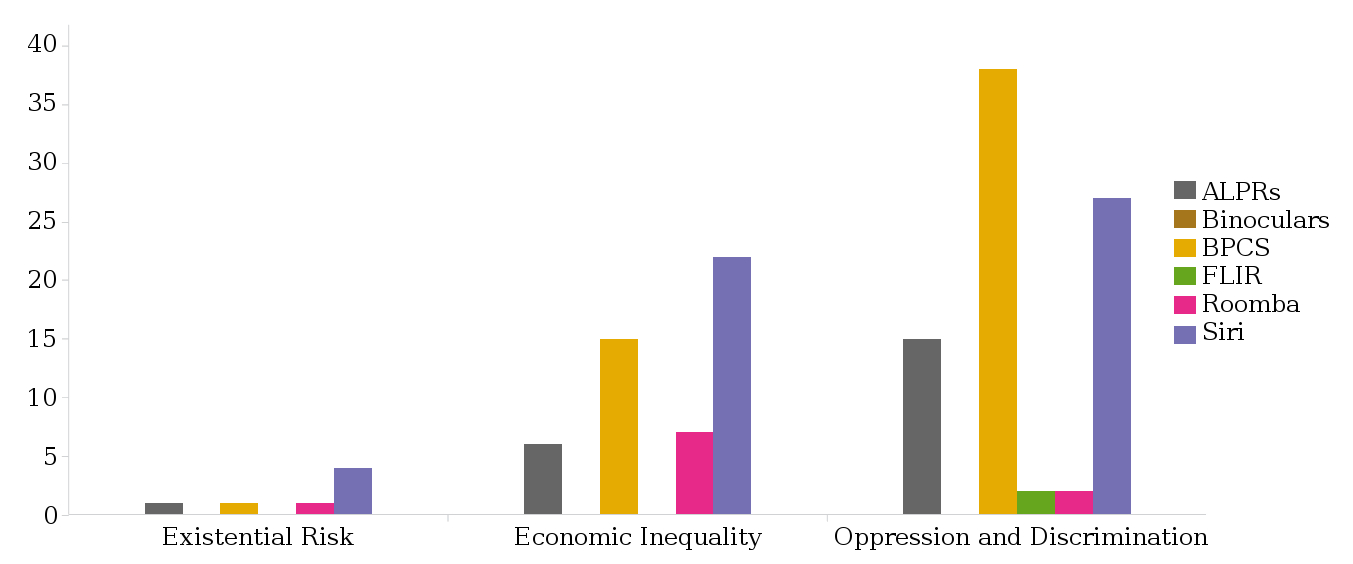}
    \caption{Frequency of survey responses indicating relevance of an AI technology to a particular issue.}
    \label{fig:tech-issues}
\end{figure}

Our results show that there is a large degree of agreement that economic inequality (82\% agree) and oppression and discrimination (83\% agree) are relevant issues, both of which are commonly associated with existing technologies.  There was more disagreement about whether existential threats are relevant (42\% agree), which is an issue more relevant to (hypothetical) human-like AI.
To further connect definitions of AI with issues about AI, we also asked a follow-up question to those participants who had rated any of the technologies we presented as AI and rated any of the issues we had presented as relevant: ``Do any of the technologies you classified as A.I., or others that come to mind, relate to the issues you indicated were relevant to the topic of A.I.?''  Figure \ref{fig:tech-issues} shows that substantial fractions of our sample considered Booking Photo Comparison Software and Siri to be relevant to oppression and discrimination, and a somewhat smaller fraction also considered those technologies to be relevant to economic inequality.  Several AI researchers also rated Automated License Plate Readers as related to these two issues.  These results show that many AI researchers consider existing, widely used information technologies to be AI and to be relevant to issues pertaining to the regulation of AI.

\section{Discussion}

The motivation of this research was to describe how AI researchers and policymakers define AI and to examine possible gaps between their definitions. A disconnect between research and policymaker definitions may lead to harmful and unintended consequences in the realm of policymaking. Public concern about AI regarding issues such as economic inequality and existential risk call for policymakers to address these issues, but the absence of useful policy-facing definitions of AI hampers the development of appropriate and effective regulatory policies.


We found that relative to AI researchers, policy documents tend to favor human thinking/behavior in their definitions of AI. One possible consequence of this is that policy documents may overly focus on the future social consequences of AI. As noted by Cave et al., \cite{cave2019scary,cave2018ai}, policymaker conceptions of humanoid AI may focus on the retreating horizon of systems still-to-be-created at the risk of passing over autonomous systems already in place.

To address this gap between policy-makers' and AI researchers' definitions of AI, and to practically address the social impacts of AI, we suggest using a definition that would maintain high fidelity to researcher definitions of AI while better lending itself to policy implementation. First, a policy-facing definition ought to fit with existing scholarly work on AI, and thus with AI researchers' conceptions of AI. Second, this definition should reflect the concerns of citizens. Such a definition should capture, rather than leave out, AI technologies of worry to the general public. This would also mean pulling back the AI conversation from future-focused issues to focus on more immediate issues.
Therefore, we propose the following necessary criteria for a policy-facing definition: (i) inclusivity of both currently deployed AI technologies and future applications, (ii) accessibility for non-expert audiences, and (iii) allowing for policy implementation of reporting and oversight procedures.

We conclude our analysis by offering an example of a recent policy definition that we believe meets these criteria.  The OECD offers the following definition: ``An AI system is a machine-based system that can, for a given set of human-defined objectives, make predictions, recommendations, or decisions influencing real or virtual environments. AI systems are designed to operate with varying levels of autonomy.''\footnote{\url{https://legalinstruments.oecd.org/en/instruments/oecd-legal-0449}}  
This definition meshes well with those in scholarly work such as given by Russell and Norvig \cite{russell2016artificial} and more recently by Bryson \cite{bryson2019past} that characterize AI as systems aimed at accomplishing a particular goal by taking input from a digital or physical environment and undertaking goal-oriented action on that information by producing a decision or other output.  In line with our criteria, the OECD definition is also specific enough to include existing technologies.  For instance, a system like Automated License Plate Readers (ALPRs) makes recommendations that influence how police interact with their environment. Finally, also in line with our criteria, this definition emphasizes that the goals of AI are \textit{human} goals, and not intrinsic to the AI itself (c.f. \cite{johnson2017reframing}).

\section{Conclusion}

Conversations about AI---including what it is and how it is affecting society---abound in policy discussions, technical research communities, and the public arena. Here, we highlighted differences between how AI researchers define AI technologies and how policy documents---including from organizations dedicated to the policy and ethics of AI---define AI technologies. 
We find that while AI researchers tend to define AI in terms of ideal thinking/behavior, policy documents tend to define AI in terms of human thinking/behavior.
An important consideration in this comparison is that definitions adhering closely to the technical functionality of AI systems are more inclusive of technologies in use today, whereas definitions that emphasize human-like capabilities are most applicable to hypothetical future technologies.  Therefore, regulatory efforts that emphasize AI as human-like may risk overemphasizing concern about future technologies at the expense of pressing issues with existing deployed technologies.

\section*{Acknowledgments}
Special thanks to many friends and colleagues for piloting, publicizing, and participating in our survey, and to Emma Spiro, Ceilyn Boyd, and danah boyd for helpful comments on our thoughts and methods.
This work was supported in part by the Washington Research Foundation, and by a Data Science Environments project award from the Gordon and Betty Moore Foundation (Award \#2013-10-29) and the Alfred P. Sloan Foundation (Award \#3835) to the University of Washington eScience Institute. Any opinions, findings, and conclusions or recommendations expressed in this material are those of the authors and do not necessarily reflect those of the sponsors.

\bibliographystyle{ACM-Reference-Format}
\bibliography{bib}

\renewcommand{\thesection}{S\arabic{section}}
\renewcommand{\thefigure}{S\arabic{figure}}
\renewcommand{\thetable}{S\arabic{table}}

\setcounter{section}{0}
\setcounter{figure}{0}
\setcounter{table}{0}

\newpage




\section{Supplementary Analysis}
\label{sec:s_analysis}

\subsection{Twitter Data Analysis of AI Issues}

In three weeks of April-May 2019,  we collected data from the social media site, Twitter, in order to confirm specific issues of concern to the general public regarding the downstream social consequences of AI.  We collected 300 tweets related to AI by using the Twitter Streaming API to search for the keywords ``AI'' and ``artificial intelligence''.  To inform our survey results, one author manually coded these tweets as related or not to existential risk, economic inequality, and oppression and discrimination. A second author validated a random sampling of these codes.
To confirm that the three issues we asked researchers about in our survey are issues of public concern, we analyzed the Twitter data we collected.  We found that out of the 300 tweets we collected, 172 or roughly 60\% were actually related to artificial intelligence (compared to e.g. ``Allen Iverson'').  Among the 172 relevant tweets, 15\% were related to at least one of the three issues, and around 5\% were related to each of the three issues we identified (8 related to existential risk, 6 to economic inequality, and 7 related to oppression and discrimination).  Given that around a total of 25\% of the tweets were related to positive or neutral issues, such as modernization, this sample shows that at the time of collection there was significant public concern about these issues.

\begin{figure}
    \centering
    \includegraphics[width=0.32\linewidth]{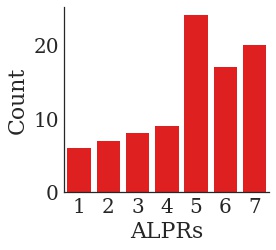}
    \includegraphics[width=0.32\linewidth]{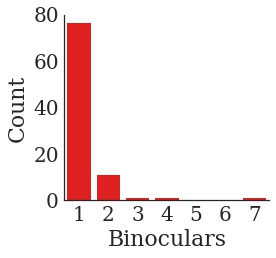}
    \includegraphics[width=0.32\linewidth]{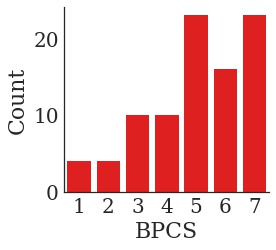}\\
    \includegraphics[width=0.32\linewidth]{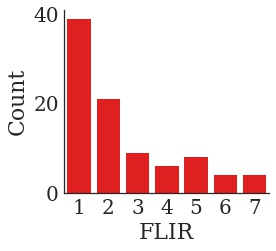}
    \includegraphics[width=0.32\linewidth]{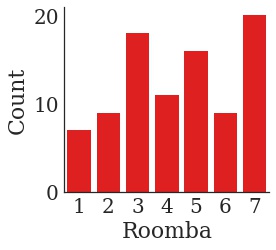}
    \includegraphics[width=0.32\linewidth]{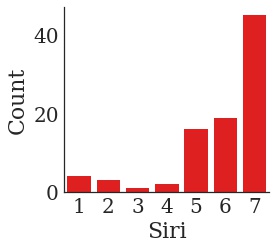}
    \caption{Histograms of AI researcher ratings of technologies as relying on AI. (1 = Strongly disagree, 7 = Strong agree)}
  \label{fig:supp-1}
  \end{figure}

\begin{figure}
  \begin{center}
    \includegraphics[width=0.2\textwidth]{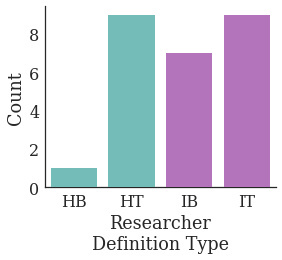}
  \end{center}
  \caption{Histogram of types of definitions AI researchers gave in our snowball sample replication of our first survey, which favor ideal (rational) behavior/thinking definitions over human-imitating behavior/thinking. Codes as in Figure \ref{fig:survey-1}.}
\end{figure}

\begin{figure}
  \begin{center}
    \includegraphics[width=0.2\textwidth]{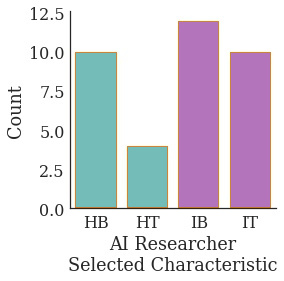}
    \includegraphics[width=0.2\textwidth]{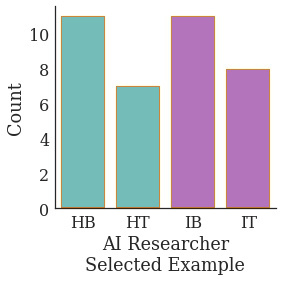}
  \end{center}
  \caption{Histogram of types of definitions AI researchers selected as favored characteristics and best example definitions in our snowball sample of our second survey. Codes as in Figure \ref{fig:survey-1}.}
\end{figure}

\begin{figure}
    \centering
    \includegraphics[width=0.32\linewidth]{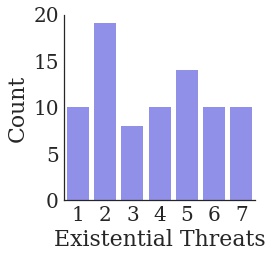}
    \includegraphics[width=0.32\linewidth]{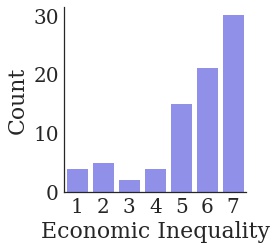}
    \includegraphics[width=0.32\linewidth]{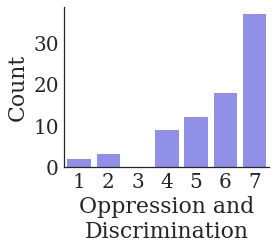}
      \caption{Histograms of AI researcher ratings of issues as relevant to AI in our snowball sample. (1 = Strongly disagree, 7 = Strong agree)}
\end{figure}

    \begin{figure}
    \centering
    \includegraphics[width=\linewidth]{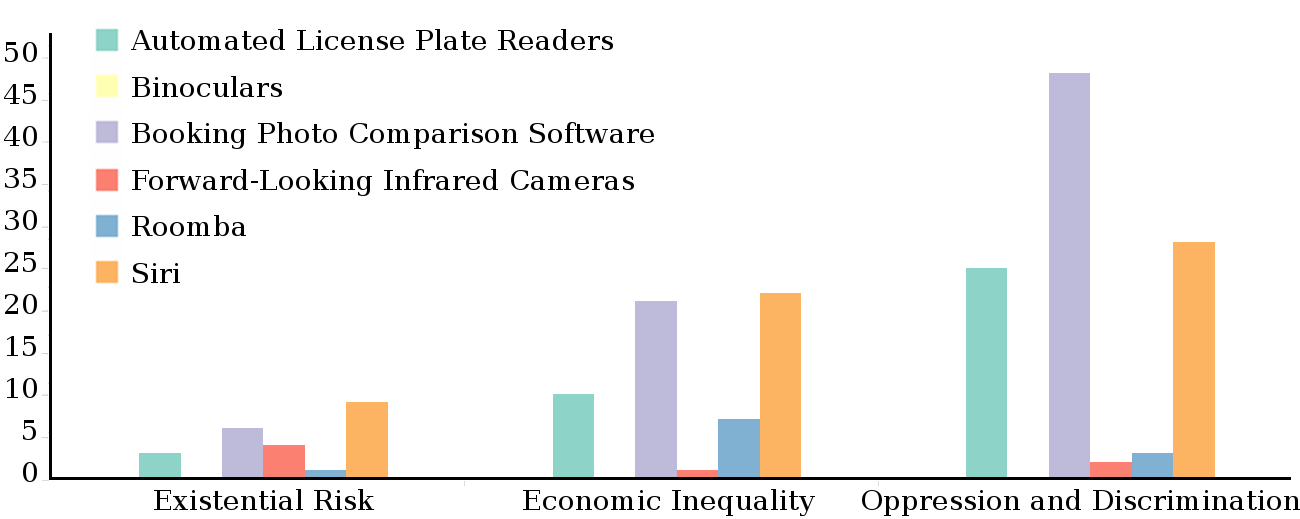}
    \caption{Histograms of which AI technologies (as chosen within each survey response) relate to each issue (as indicated relevant by the same respondent) from our snowball sample.}
    \label{fig:supp-end}
\end{figure}

\subsection{Results from Snowball Sampling Method}

Because of our relatively small sample of AI researchers through our primary survey recruitment method, we aimed to replicate our results with a snowball sampling methodology.
We collected snowball sampling data from posts on social media, emails to colleagues, and through prompts at the end of our surveys with a link to reshare. Survey participation was again voluntary and uncompensated. 

In our snowball sample for our first survey, 488 people opened a survey link, and 262 participants completed the survey. We use partial data from all participants who consented.  In our analysis we subsampled our data to only include participants self-identifying as AI researchers or publishing in at least one of five AI/ML conferences we asked about (NeurIPS, ICML, ICLR, IJCAI, or AAAI). This left a final sample size of 108 AI researchers for our analyses. In this sample, 20\% reported female as their gender, 74\% male, 3\% entered their own gender identity, and 2\% indicated they preferred not to say, with 1\% identifying as American Indian or Alaska Native, 30\% as Asian, 4\% Black or African American, 9\% Hispanic, 0\% Native Hawaiian or Pacific Islander, 47\% White, 4\% entering their own race or ethnicity, and 4\% indicating they preferred not to say.

In our snowball sample for our second survey, 136 people opened the survey link, and 74 participants completed the survey. We use partial data from all participants who consented.  In our analysis we again subsampled our data to only include participants self-identifying as AI researchers or publishing in at least one of five AI/ML conferences we asked about (NeurIPS, ICML, ICLR, IJCAI, or AAAI). This left a final sample size of 44 AI researchers for our analyses. In this sample, 22\% reported woman as their gender, 76\% man, 0\% non-binary, 0\% entered their own gender identity, and 3\% indicated they preferred not to say, with 0\% identifying as American Indian or Alaska Native, 18\% as Asian, 3\% Black or African American, 0\% Hispanic, 0\% Native Hawaiian or Pacific Islander, 62\% White, 8\% entering their own race or ethnicity, and 10\% indicating they preferred not to say.

The results, mirroring the figures in our main text, from these two snowball samples are given in Figures \ref{fig:supp-1}-\ref{fig:supp-end}.

\section{Stimuli}

\begin{figure}[htp]
    \centering
    \includegraphics[width=0.6\linewidth]{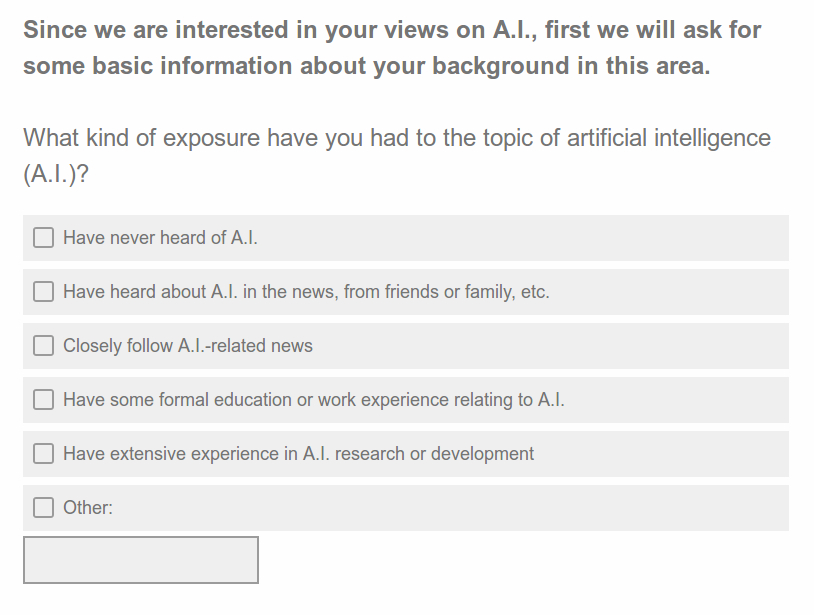}
    \caption{Images of the stimuli for AI researcher filter questions.}
    \label{fig:stimuli-1}
\end{figure}

\begin{figure}[htp]
    \centering
    \includegraphics[width=0.6\linewidth]{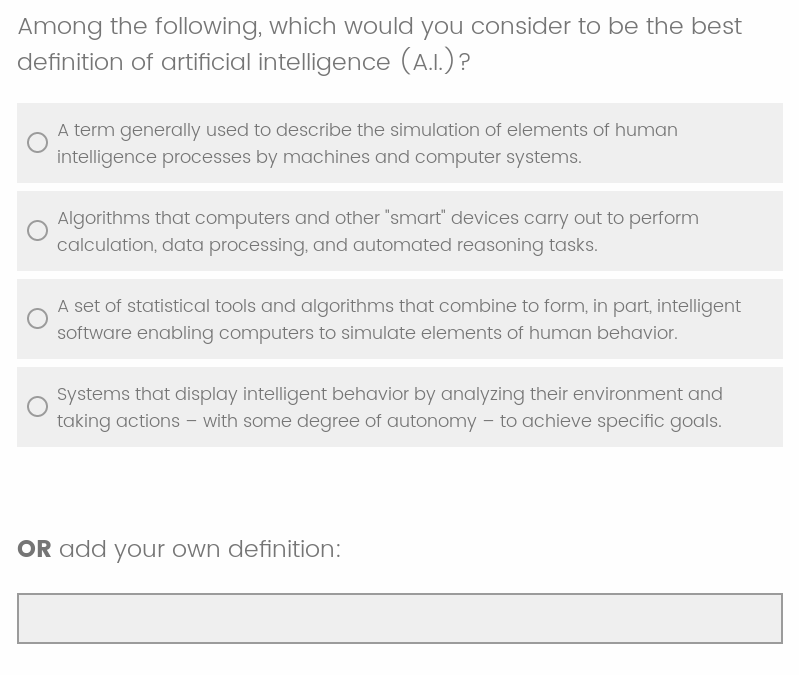}
    \caption{Image of the stimuli for our definition example question.}
    \label{fig:stimuli-end}
\end{figure}

Images of questions from our surveys are given in Figures \ref{fig:stimuli-1}-\ref{fig:stimuli-end}.

\begin{figure}[htp]
    \centering
    \includegraphics[width=0.6\linewidth]{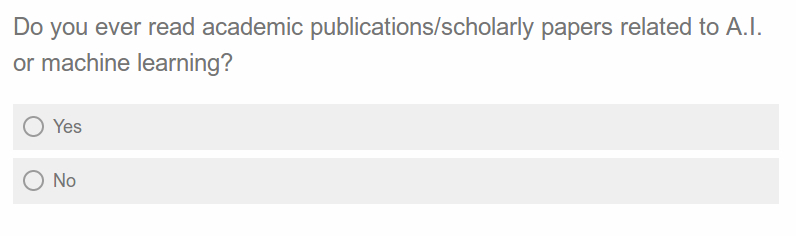}
    \caption{Images of the stimuli for AI researcher filter questions.}
    \label{fig:stimuli}
\end{figure}

\begin{figure}[htp]
    \centering
    \includegraphics[width=0.6\linewidth]{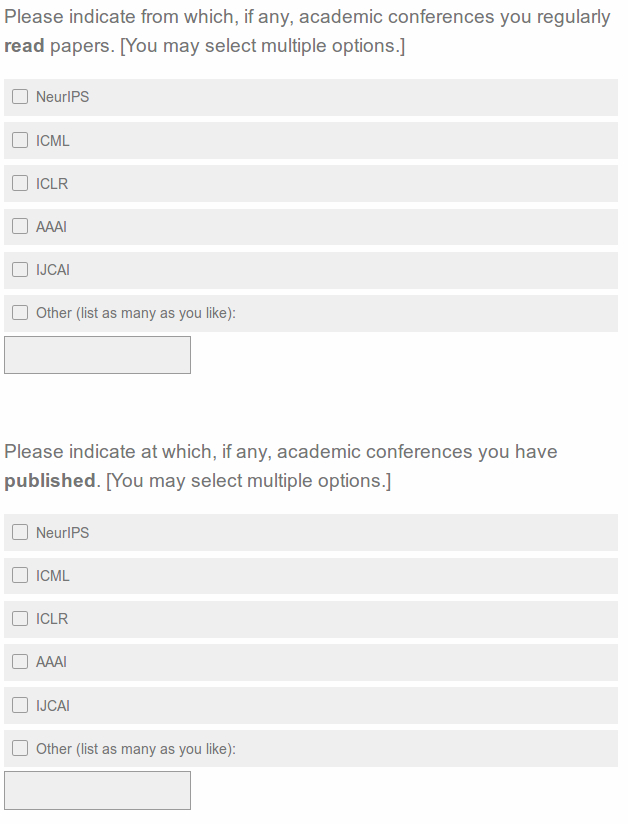}
    \caption{Images of the stimuli for AI researcher filter questions.}
    \label{fig:stimuli}
\end{figure}

\begin{figure}[htp]
    \centering
    \includegraphics[width=0.6\linewidth]{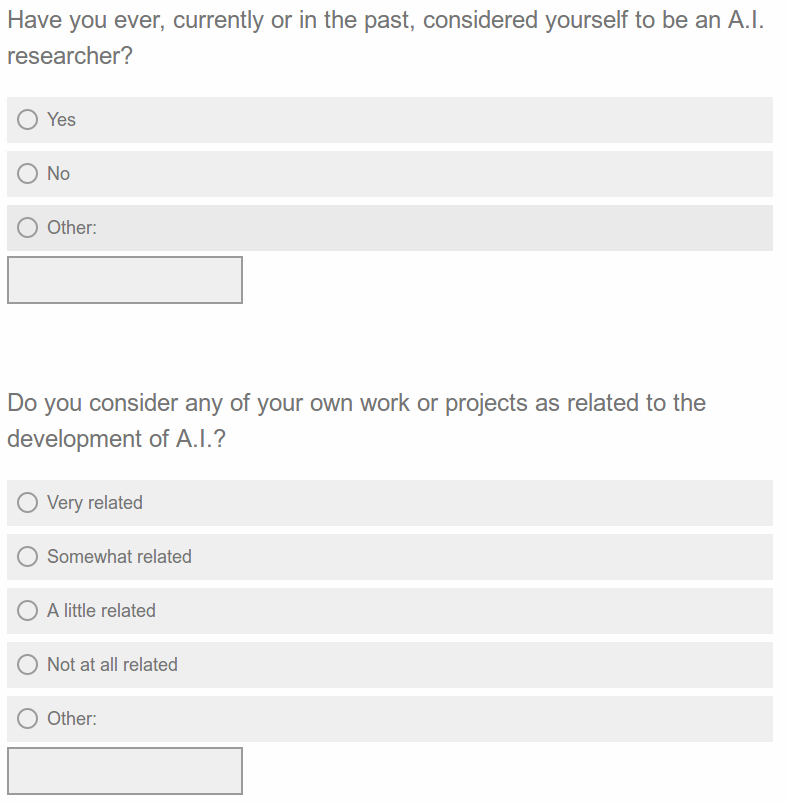}
    \caption{Images of the stimuli for AI reearcher filter questions.}
    \label{fig:stimuli}
\end{figure}

\begin{figure}[htp]
    \centering
    \includegraphics[width=0.6\linewidth]{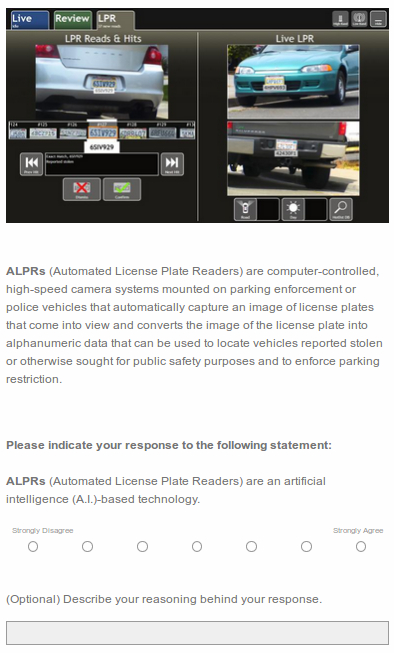}
    \caption{Images of the stimuli for our first survey's technology classification questions.}
    \label{fig:stimuli}
\end{figure}

\begin{figure}[htp]
    \centering
    \includegraphics[width=0.75\linewidth]{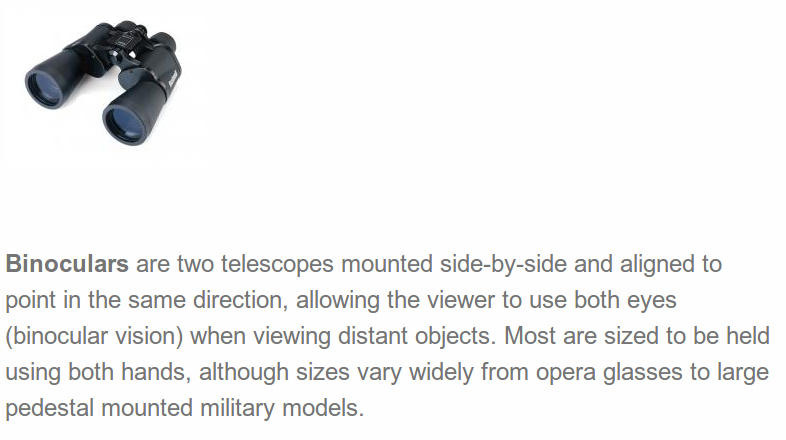}
    \caption{Images of the stimuli for our first survey's technology classification questions.}
    \label{fig:stimuli}
\end{figure}

\begin{figure}[htp]
    \centering
    \includegraphics[width=0.75\linewidth]{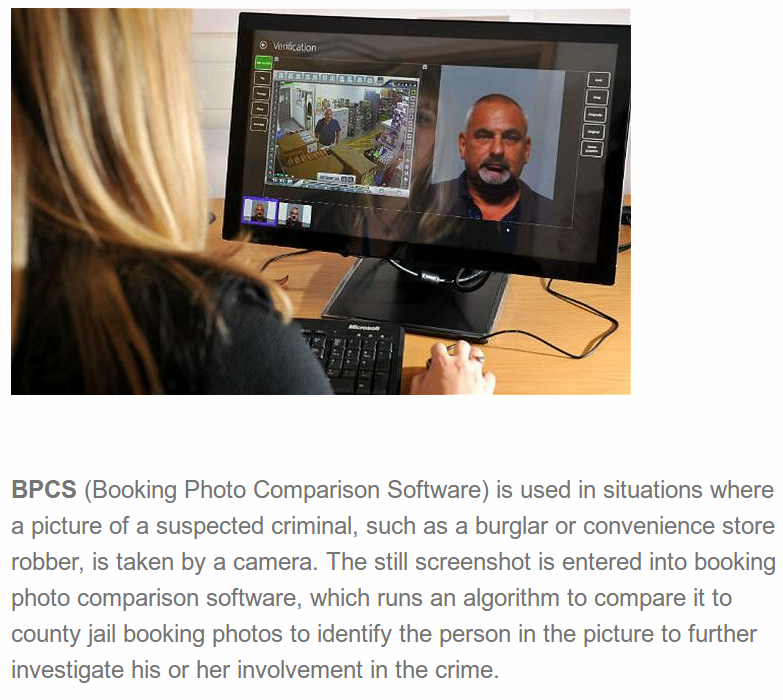}
    \caption{Images of the stimuli for our first survey's technology classification questions.}
    \label{fig:stimuli}
\end{figure}

\begin{figure}[htp]
    \centering
    \includegraphics[width=0.75\linewidth]{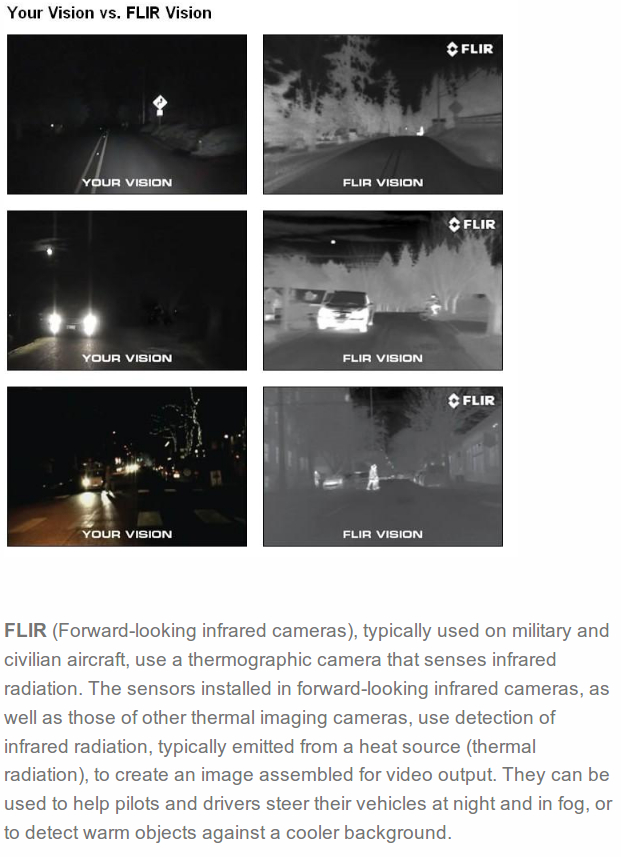}
    \caption{Images of the stimuli for our first survey's technology classification questions.}
    \label{fig:stimuli}
\end{figure}

\begin{figure}[htp]
    \centering
    \includegraphics[width=0.75\linewidth]{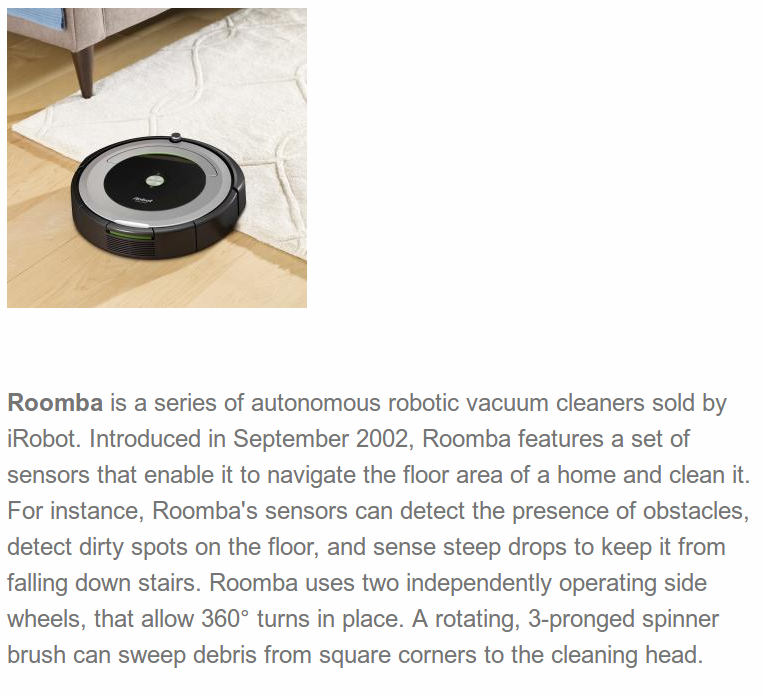}
    \caption{Images of the stimuli for our first survey's technology classification questions.}
    \label{fig:stimuli}
\end{figure}

\begin{figure}[htp]
    \centering
    \includegraphics[width=0.75\linewidth]{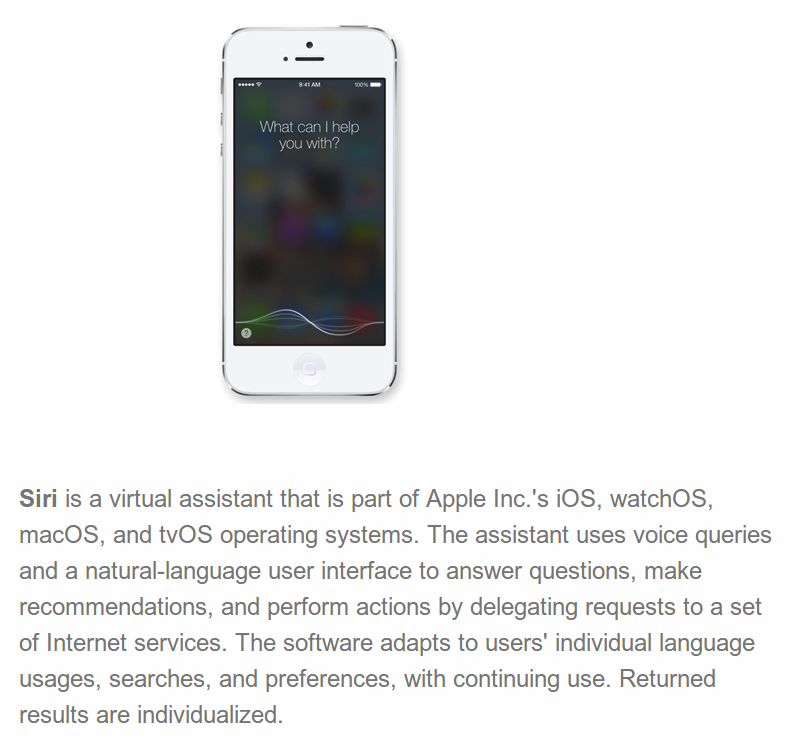}
    \caption{Images of the stimuli for our first survey's technology classification questions.}
    \label{fig:stimuli}
\end{figure}

\begin{figure}[htp]
    \centering
    \includegraphics[width=0.6\linewidth]{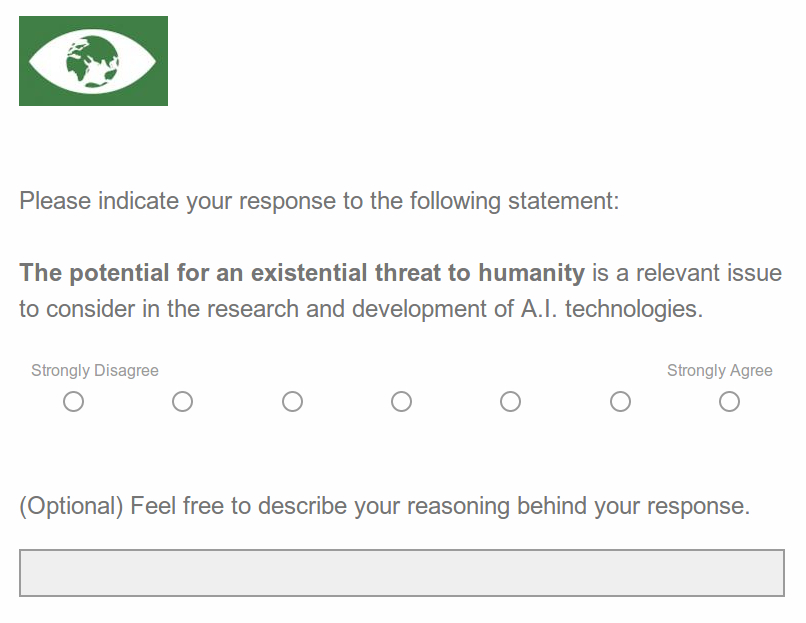}
    \caption{Images of the stimuli for our first survey's issue questions.}
    \label{fig:stimuli}
\end{figure}

\begin{figure}[htp]
    \centering
    \includegraphics[width=0.6\linewidth]{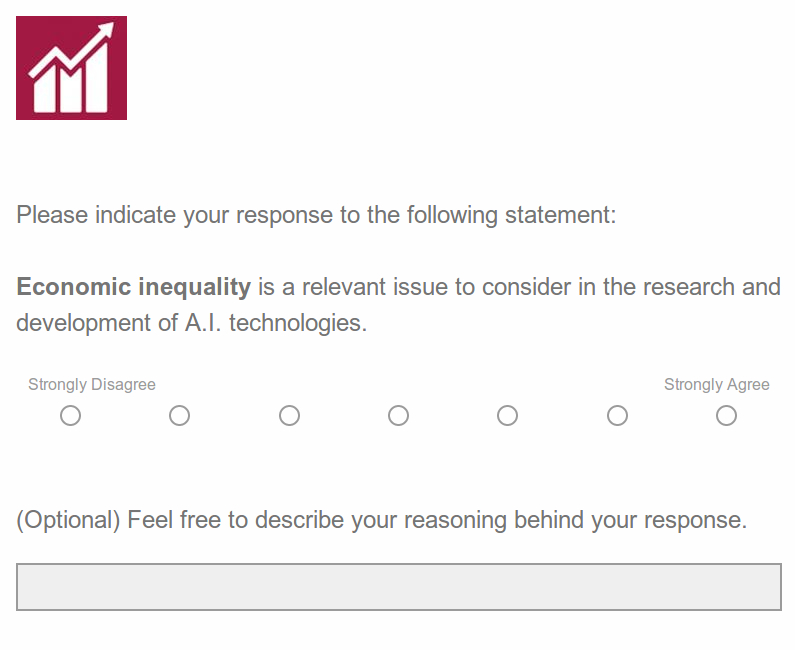}
    \caption{Images of the stimuli for our first survey's issue questions.}
    \label{fig:stimuli}
\end{figure}

\begin{figure}[htp]
    \centering
    \includegraphics[width=0.6\linewidth]{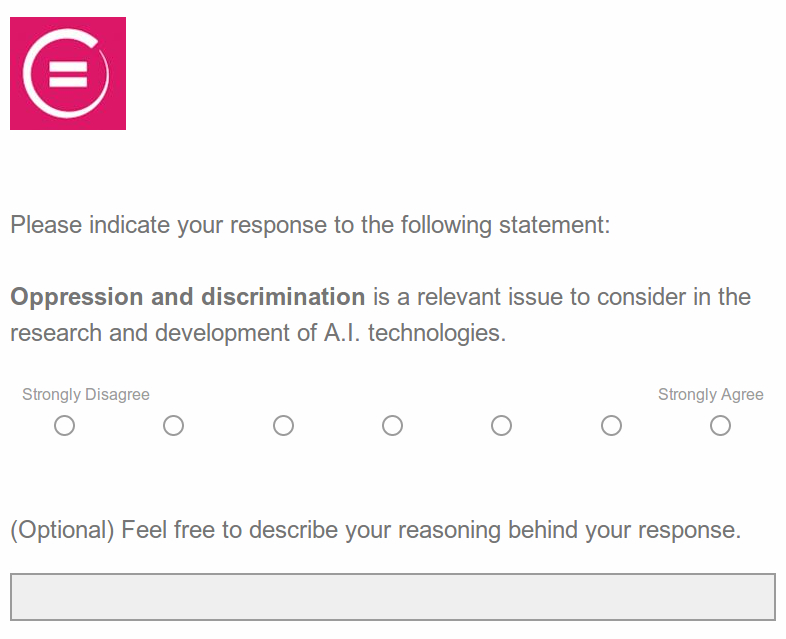}
    \caption{Images of the stimuli for our first survey's issue questions.}
    \label{fig:stimuli}
\end{figure}

\begin{figure}[htp]
    \centering
    \includegraphics[width=0.6\linewidth]{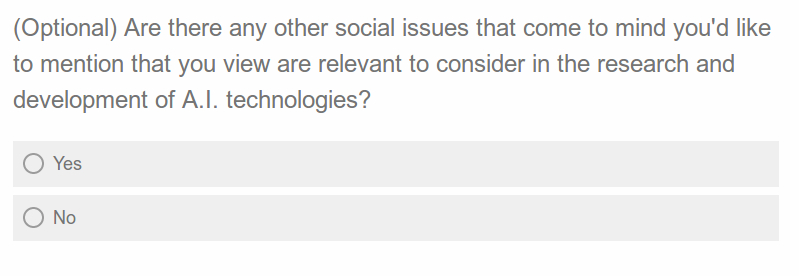}
    \includegraphics[width=0.6\linewidth]{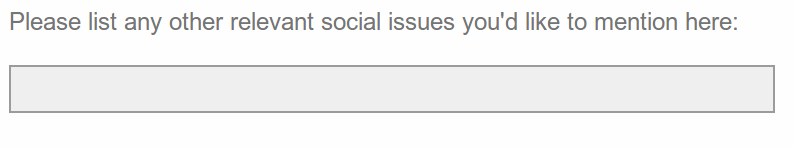}
    \caption{Images of the stimuli for our first survey's issue questions.}
    \label{fig:stimuli}
\end{figure}

\begin{figure}[htp]
    \centering
    \includegraphics[width=0.6\linewidth]{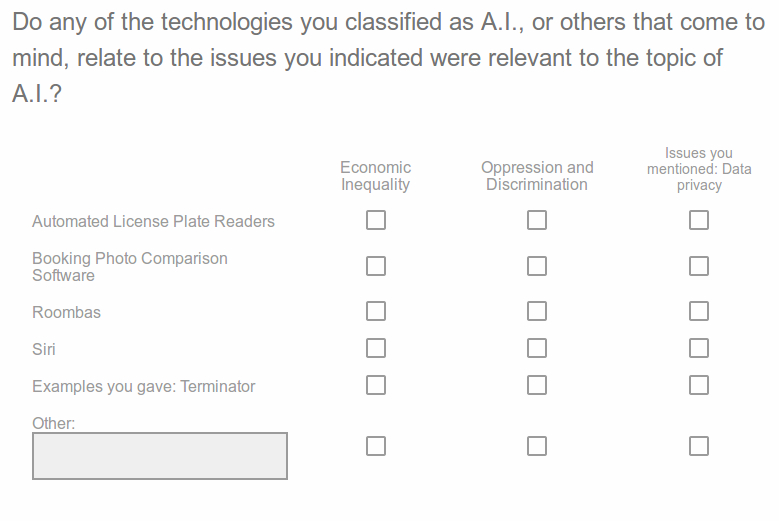}
    \caption{Images of the stimuli for our first survey's issue questions.}
    \label{fig:stimuli}
\end{figure}

\begin{figure}[htp]
    \centering
    \includegraphics[width=0.6\linewidth]{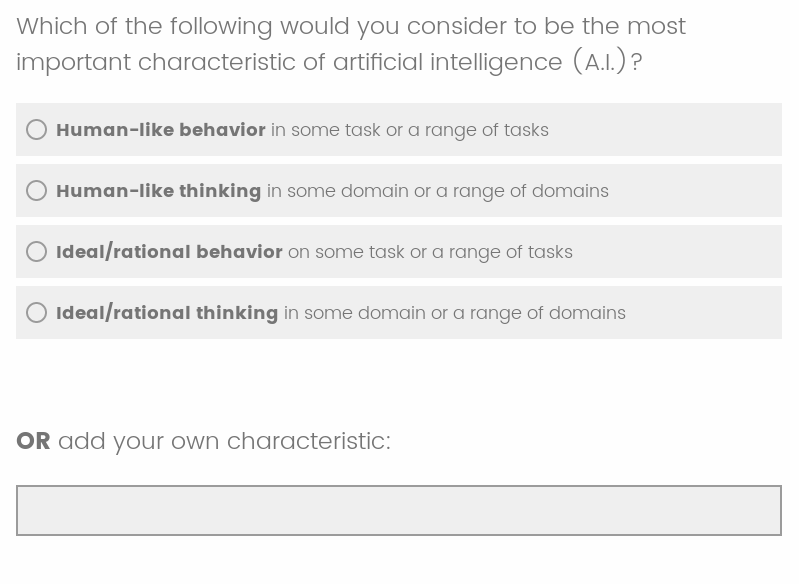}
    \caption{Image of the stimuli for our second survey's definition characteristics question.}
    \label{fig:stimuli}
\end{figure}

\end{document}